# Action potentials induce biomagnetic fields in Venus flytrap plants

Anne Fabricant[1]*, Geoffrey Z. Iwata[1], Sönke Scherzer[2], Lykourgos Bougas[1], Katharina Rolfs[3], Anna Jodko-Władzińska[3,4], Jens Voigt[3], Rainer Hedrich[2], and Dmitry Budker[1,5]

**Upon stimulation, plants elicit electrical signals that can travel within a cellular network analogous to the animal nervous system. It is well-known that in the human brain, voltage changes in certain regions result from concerted electrical activity which, in the form of action potentials (APs), travels within nerve-cell arrays. Electrophysiological techniques like electroencephalography[1], magnetoencephalography[2], and magnetic resonance imaging[3,4] are used to record this activity and to diagnose disorders. In the plant kingdom, two types of electrical signals are observed: all-or-nothing APs of similar amplitudes to those seen in humans and animals, and slow-wave potentials of smaller amplitudes. Sharp APs appear restricted to unique plant species like the "sensitive plant", *Mimosa pudica*, and the carnivorous Venus flytrap, *Dionaea muscipula*[5,6]. Here we ask the question, is electrical activity in the Venus flytrap accompanied by distinct magnetic signals? Using atomic optically pumped magnetometers[7,8], biomagnetism in AP-firing traps of the carnivorous plant was recorded. APs were induced by heat stimulation, and the thermal properties of ion channels underlying the AP were studied. The measured magnetic signals exhibit similar temporal behavior and shape to the fast de- and repolarization AP phases. Our findings pave the way to understanding the molecular basis of biomagnetism, which might be used to improve magnetometer-based noninvasive diagnostics of plant stress and disease.**

Electrophysiological measurements enable investigation of the plant signaling pathways involved in reception and transduction of external stimuli, as well as communication within the plant body. Among the stimuli which can elicit pronounced plant electrical responses are light[9], temperature[10], touch[5,11], wounding[12], and chemicals[13]. In contrast to the three-dimensional complex electrical network of the human brain, the circuitry of a plant leaf is two-dimensional only. The bilobed trap of the *Dionaea* plant (Fig. 1a,b), formed by the modified upper part of the leaf, snaps closed within a fraction of a second when touched. Three trigger hairs that serve as mechanosensors are equally spaced on each lobe. When a prey insect touches a trigger hair, an AP (Fig. 1c) is generated and travels along both trap lobes. If a second touch-induced AP is fired within 30 s, the viscoelastic energy stored in the open trap is released and the capture organ closes[14], imprisoning the animal food stock for digestion of a nutrient-rich meal. The leaf base, or petiole, is not excitable and is electrically insulated from the trap. Because of this, the trap can be isolated functionally intact from the plant by a cut through the petiole. On the isolated trap, mechanical stimuli trigger APs and closure just as on the intact plant. For the comfort of electrophysiological studies, one of the trap lobes can be fixed to a support while the other is removed, without affecting the features of AP firing. It has been shown that this simplified experimental flytrap system is well-suited to study the AP under highly reproducible conditions[15]. Other than by touch or wounding (mechanical energy), traps can be stimulated by salt loads (osmotic energy)[16] and temperature changes (thermal energy).

[1] Helmholtz Institute Mainz, GSI Helmholtzzentrum für Schwerionenforschung, Darmstadt, Germany; Johannes Gutenberg University of Mainz, Germany

[2] Department of Molecular Plant Physiology and Biophysics, University of Würzburg, Germany

[3] Physikalisch-Technische Bundesanstalt, Berlin, Germany

[4] Warsaw University of Technology, Faculty of Mechatronics, Warsaw, Poland

[5] Department of Physics, University of California, Berkeley, CA, USA

* e-mail: afabrica@uni-mainz.de

Since touch activation of APs can cause unwanted mechanical noise in electric and magnetic recordings, we use thermal stimulation in our experiments. The interdisciplinary work presented here encompasses two complementary sets of experiments: the temperature dependence of flytrap electrical activity was studied in a plant-physiology laboratory, while magnetometer measurements of heat-stimulated traps were conducted in a magnetically shielded room.

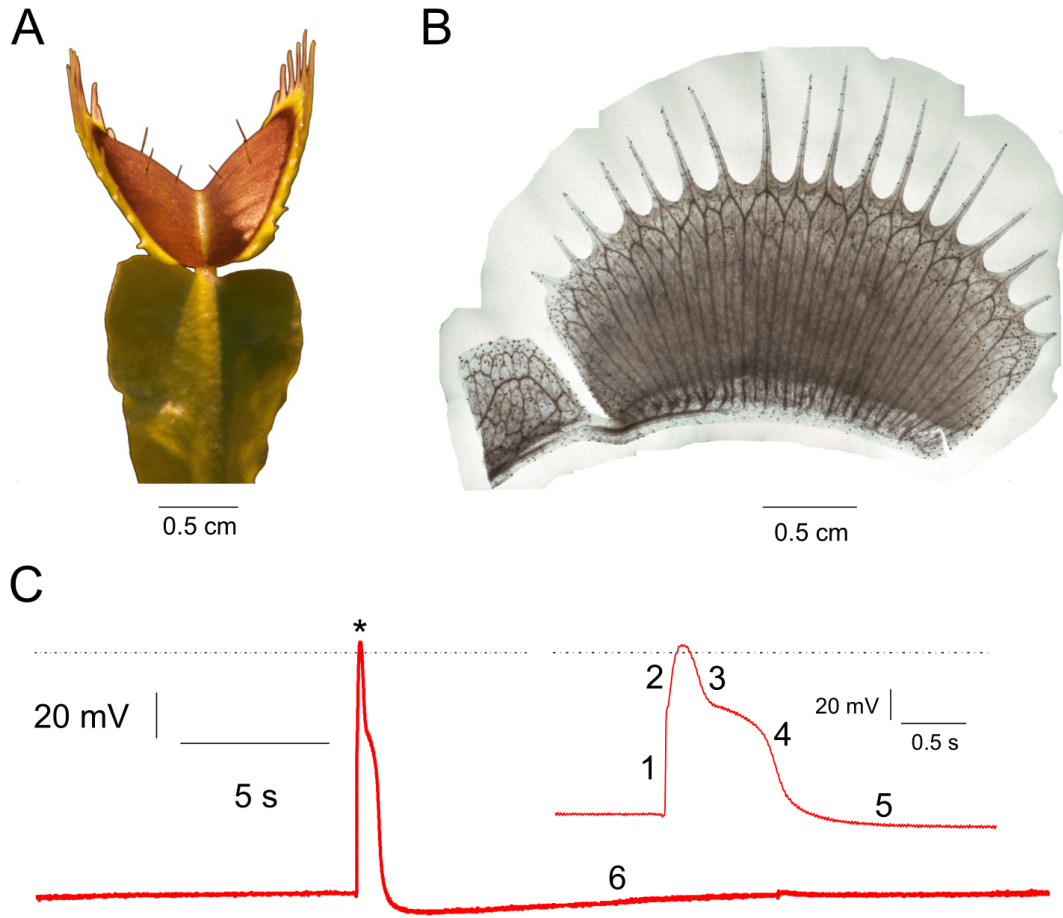

**Fig. 1 | Venus flytrap geometry and action potentials. a**, *Dionaea muscipula* leaf forms into a bivalved snap trap connected to the leaf base, or petiole. **b**, Side view of a single trap lobe showing vasculature structure. In contrast to the petiole, the trap contains parallel veins of interconnected cells. These veins consist of both dead low-conductivity water pipes (xylem) and living conductive phloem. Here the vasculature was imaged by staining for the dead vascular tissue. **c**, Intracellular AP lasting 2 s is subdivided into six phases (numbers), as explained in the text. The depolarization peak is indicated by an asterisk; the dotted line represents 0 mV. Inset, Zoom-in on the AP, resolving the first five phases of the AP.

**Heat-induced action potentials**

When we heated up the support to which excised open traps were fixed, APs were elicited and the traps closed (Extended Data Fig. 1). To study the temperature dependence of heat-induced AP initiation, on one of the trap lobes we mounted a clamp equipped with a Peltier device and surface-voltage electrode (Fig. 2a). From a resting temperature of 20°C, the trap temperature was increased monotonically to 45°C at a rate of 4°C/s (Extended Data Fig. 2). Below 30°C, no APs were observed; above 30°C, the probability of AP firing increased and was maximal (100%) above 40°C. In 60 independent experiments using 10 different traps, we recorded the temperature at which an AP was first induced. When these data were plotted as temperature-dependent AP-firing probability (Fig. 2b), the curve could be well-fitted by a single Boltzmann equation characterized by a 50% AP-firing probability at 33.8°C. This behavior indicates that heat activation of the AP is based on a two-state process. The ion channels that carry the classical animal-type AP also occupy two major states: closed and open. In contrast to the animal sodium-based AP, the plant AP depolarization is operated by a calcium-activated anion channel[5]. Thus, we conclude that the temperature "switch" of the *Dionaea* AP is based on a calcium-dependent process. Following $Ca^{2+}$ binding, the anion-channel gates open. Our experiments indicate that at temperatures of $T \lesssim 34°C$ the cellular $Ca^{2+}$ level remains below threshold, but at $T \gtrsim 34°C$ there is enough chemical energy to open a critical number of anion channels, driving the fast depolarization phase of the AP.

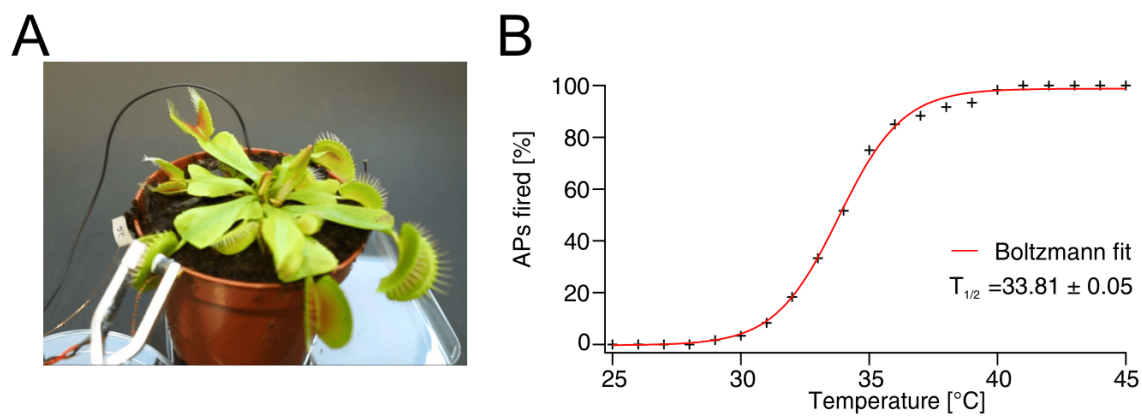

**Fig. 2 | Electrical measurements of heat-induced action potentials. a**, *Dionaea* plant with clamp mounted on one lobe of a trap, equipped with a Peltier device and surface-voltage electrode. A ground electrode is placed in the soil surrounding the plant root. **b**, Temperature dependence of AP-firing probability fitted by a Boltzmann equation (red curve), characterized by 50% firing probability at temperature $T_{1/2}$.

The *Dionaea* AP can be subdivided into 6 well-defined phases (Fig. 1c): (1) fast depolarization, (2) slow depolarization, (3) fast repolarization, (4) slow repolarization, (5) transient hyperpolarization, and (6) slow recovery of the membrane potential to the pre-AP state. When comparing APs recorded at different temperatures, we found that temperature affects the signal amplitude and duration. Increasing the thermal energy input changed not only the probability for an AP to be fired, but also led to an increased AP amplitude and decreased half-depolarization time (Supplementary Information). These facts indicate that heat-sensitive ion channels trigger and shape the AP: at higher temperatures, thermal energy input causes more closed $Ca^{2+}$-activated anion channels to open and depolarize the membrane potential. Compared to depolarization, fast repolarization (mediated by $K^+$ channels) and transient hyperpolarization (caused by depolarization activation of outward-directed protein pumps) were much less affected by temperature. The recovery time to reach the resting membrane potential was essentially insensitive to temperature changes.

Besides lowering the AP firing threshold and changing certain features of the AP, prolonged heat stimulation can induce trap lobes to enter an autonomous AP firing mode (Extended Data Fig. 1). When increasing the bottom surface temperature of the recording-chamber base from 20 to 46°C, AP spiking activity sets in after a couple of seconds, reaching a steady AP firing frequency of 3.8 per minute at a stable 46°C surface temperature. Induction of autonomous APs has also been obtained using flytraps treated with NaCl salt (osmotic energy)[17].

**Biomagnetism**

Having established heat stimulation as a reliable noninvasive technique for inducing flytrap APs, we searched for the magnetic field associated with this electrical excitability. Magnetometry experiments were carried out at Physikalisch-Technische Bundesanstalt (PTB) Berlin in the Berlin Magnetically Shielded Room 2 (BMSR-2) facility[18], using four QuSpin Zero-Field Magnetometers (QZFM). These commercial optically pumped magnetometers (OPMs) employ a glass cell containing alkali vapor to sense changes in the local magnetic-field environment[8]. A magnetically shielded environment is required for operation of the magnetometers, and use of a walk-in shielded room allowed for the constant presence of an experimenter to prepare plant samples and carry out measurements. As shown in Fig. 3, an isolated trap lobe was attached to the housing of the primary sensor (denoted A), such that the distance between the plant sample and the center of the atomic sensing volume was approximately 7 mm. Two secondary sensors (B and C) were placed nearby the primary sensor to measure signal fall-off, and an additional background sensor (D) was used to monitor the magnetic environment in the shielded room. Each magnetometer is sensitive to signals along two orthogonal axes. Sensor electronics were connected to a data-acquisition system in the PTB control room outside the magnetically shielded room. To monitor heat-induced APs, we used two silver-tipped copper surface electrodes, inserted in either end of the plant sample[19]. These data, together with other auxiliary trigger signals, were sampled simultaneously with the OPMs using the same data-acquisition system.

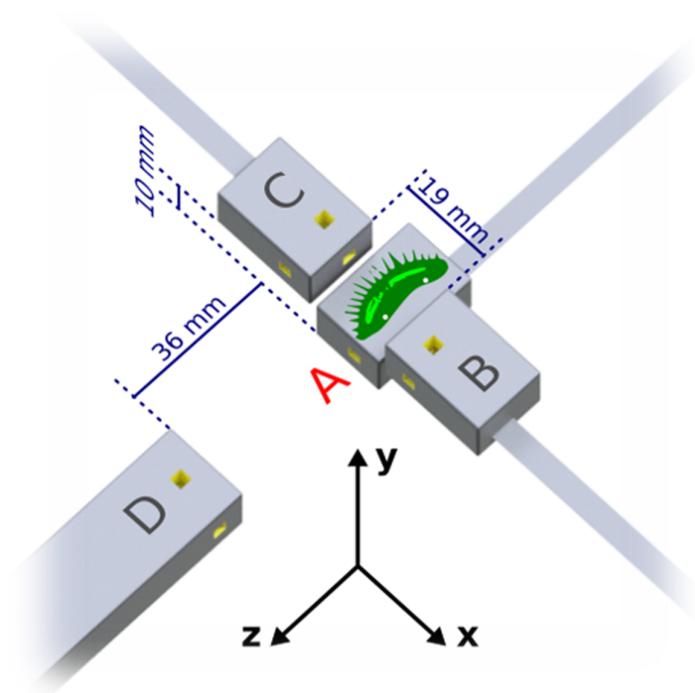

**Fig. 3 | Schematic of the experimental setup in the magnetically shielded room.** The plant sample, an isolated lobe of the flytrap, is placed on top of primary sensor A, in the *x-z* plane with trigger hairs exposed. For reference, the dimensions of the housing (gray boxes) for the primary and secondary sensors are 24.4×16.6×12.4 mm$^3$. Yellow cut-outs indicate the position of the 3×3×3 mm$^3$ atomic sensing volume. A 3D-printed ABS plastic structure (not shown) holds the magnetometers in position on a wooden table. White dots on the plant sample, approximately 1 cm apart, indicate the placement of the surface electrodes for AP monitoring. In the coordinate system shown, all magnetometers are sensitive along the *y*-axis, normal to the surface of the plant sample; furthermore, A and D are sensitive along the *z*-axis, and B and C are sensitive along the *x*-axis. Sensors B and C are positioned symmetrically around sensor A. Sensor D serves as a background sensor and is therefore located farther away from the sample.

Resistive heaters in the magnetometer housing, which are used to increase the atomic density and improve sensitivity, also served to induce autonomous AP firing via surface heat transfer. Electric and magnetic signals were recorded simultaneously from traps heated to a surface temperature of 41°C. Prior to the measurements, we performed tests to ensure that no spurious magnetic fields were generated by the electrode system (Supplementary Information). To better distinguish the observed magnetic signals from background noise, we triggered on the electric signals and averaged the magnetic data in a time window around those trigger points. Examples of averaged magnetic data are shown in Fig. 4a,b. A clear magnetic signal with a time scale corresponding to that of the averaged electric signal is visible in the primary-sensor data. For comparison, data from several different experiments were plotted (Fig. 5). To minimize common background noise, we subtracted the magnetic data of sensor D to create a gradiometer with a 48-mm baseline. Signals of up to 0.5 pT are visible in the *y*-axis gradiometric data, normal to the sample surface. The signal magnitude obtained was comparable to what one observes in surface measurements of nerve impulses in animals[20].

To quantify the significance of the signals, signal-to-noise ratios (SNRs) were calculated from the average *y*-axis gradiometric time traces in Fig. 5 as follows. The noise level is defined as the standard deviation of the gradiometric response in a 1.5 s time window (from time $t = -2$ s to $t = -0.5$ s in Fig. 4a) prior to signal onset. The signal size is defined as the amplitude of the extreme (minimum) field value, with respect to the mean value in the noise window. For the four experiments shown in Fig. 5, the SNRs range from 8 to 20. The corresponding p-values are $p < 9 \times 10^{-16}$, indicating that the probability of such signals arising from random noise is negligible. At the sub-Hz signal frequency, the sensitivity of the gradiometer is approximately 100 fT/√Hz (Extended Data Fig. 6). For both the electric and magnetic signals, the full width at half extremum (maximum or minimum, FWHM) were also calculated, where the extremum is defined with respect to the mean value in the noise window.

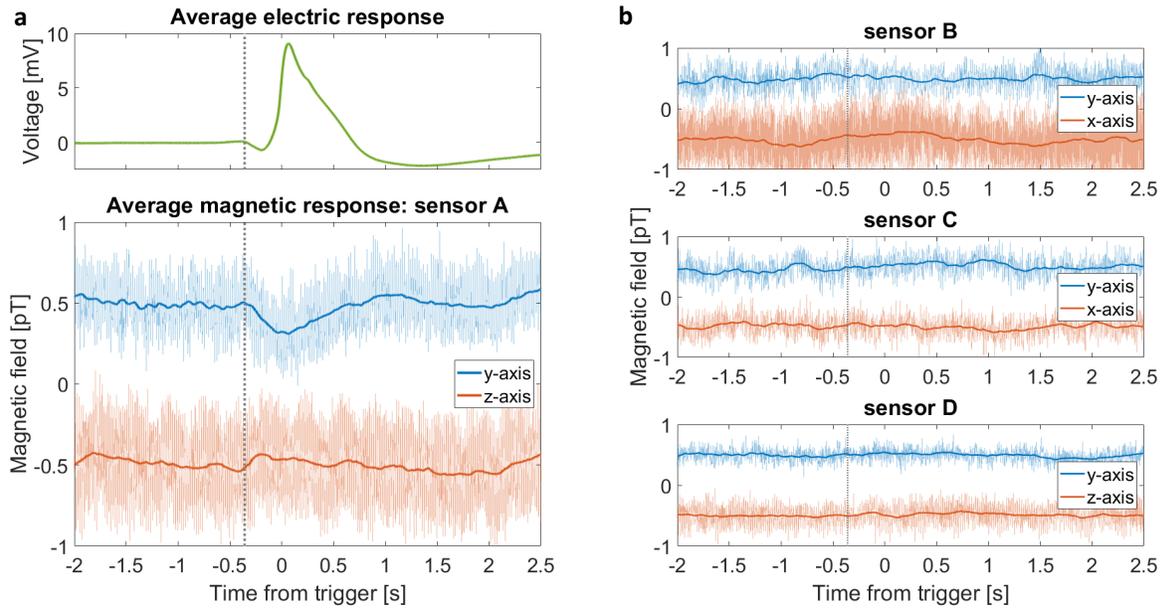

**Fig. 4 | Average action potential and corresponding magnetic signals. a**, Result of triggering on nine consecutive APs from a trap lobe heated to 41°C, then averaging the electric and magnetic data from a 4.5 s window around each trigger point. The average magnetic traces (bottom graph, opaque traces) were frequency-filtered (50 Hz low-pass), then smoothed with a 0.2 s running average. A magnetic signal is visible in both sensitive axes of the primary sensor A. For comparison, the raw unfiltered data are plotted behind the processed data. For visual clarity, DC offsets have been added to the data, and vertical gray dotted lines indicate the approximate start time of the electric signal. **b**, Average magnetic response from the other three sensors,

obtained using the same procedure as in **a**. The data from the secondary sensors, B and C, do not show a signal. The data from the background sensor D can be used to remove noise common to all sensors (see Fig. 4).

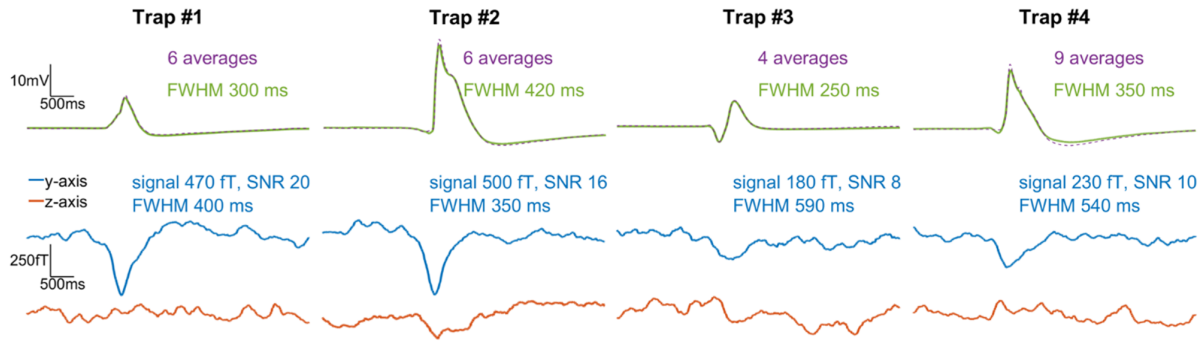

**Fig. 5 | Comparison of average electric and gradiometric signals from four different experiments.** In each case we triggered on heat-induced APs in the electric trace and performed the same data analysis as for Fig. 4. The electric response recorded by the surface electrodes (top row; average signal plotted as solid green, single AP plotted as dashed purple) varies in amplitude because a different plant sample was used in each experiment. To produce the gradiometric plots (bottom graphs) we subtracted the magnetic data of background sensor D from that of primary sensor A. The number of averages in each experiment is indicated, along with the amplitude and signal-to-noise ratio (SNR) of the *y*-axis gradiometric signal. The rightmost panel (Trap #4) shows the same data set as in Fig. 4.

The temporal superposition of the electric and magnetic signals in Fig. 5 suggests that we have detected the magnetic activity associated with the flytrap AP. Unlike in measurements of animal nerve axons and the large internodal cells of *Chara corallina* alga, where the magnetic field is proportional to the time derivative of the intracellular voltage[21,22], the magnetic signal from the complex multicellular flytrap lobe has a shape similar to the electric signal. We see features in the magnetic signal which appear to correspond to the depolarization and repolarization phases of the AP. In electric recordings using surface electrodes, the exact shape and duration of signals are dependent on the placement of electrodes on the measured sample. By contrast, magnetometry records a "true" physical signal from the organism. In this sense, it is comparable to intracellular electrode techniques. Whereas intracellular electrodes are sensitive to electrical activity of single cells, magnetometers can record both local and systemic activity at the multicellular level.

The physical origin of the measured biomagnetic fields is related to an outstanding question in plant electrophysiology: how electrical signals propagate over long distances through the plant. Essentially this is a scaling problem: while electrical signaling is well-understood in some unicellular plant systems[22], much less is known about the propagation mechanisms of such signals between cells and along cellular pathways. For the Venus flytrap system, it is known from electrode measurements that APs propagate through the trap at speeds of around 10 m/s [6]. A proposed pathway of long-distance signal propagation between plant cells in the trap is the electrically conductive phloem in the vasculature (Fig. 1b). Given that the typical resistance between two points on a trap is $R \approx 1$ MΩ [23], we can perform a basic calculation to confirm that the magnitude of the magnetic fields we measure is reasonable. We estimate the expected magnetic-field magnitude at the center of the sensing volume to be

$$B \approx \frac{\mu_0 I}{2\pi r}, \tag{1}$$

where $I = V/R \approx 10$ nA is the current passing through the trap between the electrodes, and $r \approx 7$ mm is the perpendicular distance from the trap surface. Using these values, we find $B \approx 0.3$ pT, a magnitude which corresponds well with the $y$-axis experimental results of sensor A. Although the precise distribution and directionality of current flow in the trap is unknown, we can use the geometry of the trap (Fig. 1a,b) and magnetometry setup (Fig. 3) to further interpret our results. If the $x$-oriented parallel-cable structure of the vasculature is the primary conduction pathway, magnetic field along the $y$-direction is expected at the primary sensor A, but not at the secondary sensors B and C. The symmetry of the trap about the $x$-direction could explain the relative lack of $z$-axis magnetic signal in our measurements. (Trap curvature and misalignment with respect to the sensor housing may give rise to $z$-axis signals in some experiments.) Thus, our magnetometry results agree with a hypothesis that the vasculature serves as a network for long-distance electromagnetic signaling within the trap.

**Discussion**

Although human and animal biomagnetism are well-developed areas of research[2,20,21,24,25,26], very little analogous work has been conducted in the plant kingdom[9,12,22,27], largely because biomagnetic signals are typically much smaller in amplitude and frequency than their animal counterparts. Previously reported detection of plant biomagnetism, which established the existence of measurable magnetic activity in the plant kingdom, was carried out using superconducting-quantum-interference-device (SQUID) magnetometers[9,12,22]. Atomic magnetometers are arguably more attractive for biological applications, since, unlike SQUIDs[28,29], they are non-cryogenic and can be miniaturized to optimize spatial resolution of measured biological features[20,26,30]. Our study of plant biomagnetism using atomic magnetometers documents: (i) existence and features of biomagnetic signals in the Venus flytrap, (ii) magnetic detection of APs in a multicellular plant system generally, and (iii) electric and magnetic detection of heat-induced APs in the Venus flytrap. In the future, the SNR of magnetic measurements in plants will benefit from optimizing the low-frequency stability and sensitivity of atomic magnetometers. Just as noninvasive magnetic techniques have become essential tools for medical diagnostics of the human brain and body, this noninvasive technique could also be useful in the future for crop-plant diagnostics—by measuring the electromagnetic response of plants facing such challenges as sudden temperature change, herbivore attack, and chemical exposure.

**Methods**

To obtain strong electric and magnetic signals, the health of the plants is paramount. We purchase adult Venus flytraps from a carnivorous-plant greenhouse (Gartenbau Weilbrenner, Freinsheim, Germany). Normally the plant samples are housed in a growth chamber manufactured by Poly Klima. To keep the flytraps alive during the PTB measurement run, we used homemade plastic greenhouses equipped with plant-cultivation lighting and temperature and humidity monitoring. The plants were kept on an automated 12/12-hour light/dark cycle at approximately 25°C and 75% relative humidity, treated only with distilled water.

For recording of flytrap APs in our heat-stimulation investigations, we used surface electrodes measuring the extracellular potential of a trap. One silver electrode was inserted into the trap, with the electrical connection enhanced by application of a droplet of contact gel (Laboklinika), while the reference electrode was inserted into wet soil or the petiole midrib. Electrical signals were amplified 100-fold and recorded with Patchmaster software (HEKA). Temperature dependence of AP induction was studied by application of a homemade Peltier device powered by a PTC-10 temperature-control system (npi electronic, NJ 08510, United States). Constant heat was applied using an IKA RET basic hot plate (IKA-Werke GmbH & Co. KG, Staufen, Germany) heated to 46°C.

Several types of magnetometry experiments were conducted at PTB: controls, OPM measurements using four QuSpin sensors (three Gen-2: denoted A, B, C; one Gen-1.5: denoted D), and measurements using the multi-channel SQUID array of BMSR-2. See Supplementary Information for further details of the SQUID measurements.

For the OPM measurements of isolated trap lobes, each sample was cleaved from the plant with a razor blade and placed on the primary sensor A for immediate measurement. The sample was either secured to the sensor housing with double-sided adhesive tape (acrylate, thickness 0.5 mm) or placed on a plastic slide (PET, thickness 0.22 mm) on the housing. Electrode, magnetometer, and electric reference signals were recorded at a 500-Hz acquisition rate on a 9-channel analog data-acquisition system with PC control. The raw difference signal from the two surface electrodes was first sent through a voltage preamplifier (Stanford Research Systems, Model SR560), AC-coupled with a gain of 100. It is essential to use a voltage, rather than current, preamplifier to avoid currents in the electrical leads whose magnetic fields may be detected by the magnetometers. Since leakage of electrical signals into magnetic channels is a serious concern, we address the topic in detail in Supplementary Information.

**Data availability**

The datasets generated and analyzed during the current study are available from the corresponding author on reasonable request.

**Acknowledgements**

We acknowledge the support of the Core Facility "Metrology of Ultra-Low Magnetic Fields" at Physikalisch-Technische Bundesanstalt, which receives funding from the Deutsche Forschungsgemeinschaft (DFG KO 5321/3-1 and TR 408/11-1). Dr. Tilmann Sander-Thömmes and Sophia Haude assisted during the PTB data run. Dr. Rob Roelfsema of the University of Würzburg provided valuable guidance in the early stages of the project. Pavel Fadeev offered helpful comments on the manuscript. A.F. was supported by a Carl-Zeiss-Stiftung graduate fellowship. This research was supported in part by the German Federal Ministry of Education and Research (BMBF) within the Quantumtechnologien program (FKZ 13N14439), as well as the DFG Koselleck award HE 1640/42-1 to R.H.

**Author contributions**

A.F. and D.B. proposed to study biomagnetism in the Venus flytrap. A.F., G.I., S.S., L.B., K.R., A.J.W., and J.V. conducted experiments. A.F. and S.S. analyzed data. A.F., R.H., and S.S. wrote the manuscript. R.H., D.B., G.I., L.B., S.S., J.V., A.J.W., and K.R. edited the manuscript. D.B. and R.H. supervised research.

**Competing interest declaration**

The authors declare no competing interests.

**Supplementary information**

As part of our study of heat-induced flytrap electrical behavior, we compared the amplitude and depolarization kinetics of APs recorded at 10, 20, 30, and 40°C. There was a 1.6-fold increase in AP amplitude from 10 to 40°C. When heating the trap from 10 to 30°C, the half-depolarization time dropped from $0.29 \pm 0.08$ s to $0.13 \pm 0.02$ s.

In the PTB data run, as a complement to the OPM measurements we also conducted two types of experiments using 57 channels of the BMSR-2 built-in SQUID array. The first type involved placing an intact flytrap plant directly under the SQUID dewar, whose bottom surface has a 2.8-cm offset from the plane of the pick-up coils. We closed each trap in turn by two consecutive mechanical stimulations of the trigger hairs with a plastic pipette tip. In the data analysis, we looked for signals in the magnetic data corresponding to either the APs or subsequent trap closure. Even after averaging multiple SQUID channels, no signals were found, probably because of the large distance between sample and sensors. In the second type of SQUID experiment, we attached an isolated trap lobe directly to the bottom of the dewar and performed mechanical stimulation, but again no magnetic signals were found during data analysis. Following the calculation in the main body of the paper, at offset distance of at least 2.8 cm from the sample we would reasonably expect a magnetic-field magnitude on the order of 10 fT. Since this is approximately the noise floor of the SQUID magnetometer system at 1 Hz (under ideal operating conditions), the null result is consistent with expectations.

Prior to the data run, we tested the electrode system to ensure that no spurious magnetic fields due to currents in the electrode wires would be picked up by the magnetometers under usual experimental conditions. These tests were conducted using the circuit depicted in Extended Data Fig. 3, with a four-layer MS-2 magnetic shield from Twinleaf containing two active QuSpin sensors (B and C) placed side-by-side. A function generator (Tektronix AFG2021) in parallel with a resistor created a sawtooth "artificial flytrap action potential" signal at 1.2 Hz. This signal was sent through a low-noise voltage preamplifier (SRS Model SR560) with typical experimental settings (6-dB low-pass 10-Hz filter, AC coupling, gain 1000, input impedance 100 M$\Omega$) that yielded a preamplifier output of amplitude 2 V—corresponding to the output amplitude we would see in an actual flytrap experiment. The crucial step was to simulate a "worst-case scenario" for the electrode wires. To that end, a 1-cm copper coil, in series with the preamplifier, was placed directly on top of sensor C. As is evident in Extended Data Fig. 4, no signal at 1.2 Hz was visible in the $y$-axis or $z$-axis data of either magnetometer. Even when we increased the amplitude of the electric signal by five times (corresponding to 10-V preamplifier output), no signal at 1.2 Hz was observed. Thus, we were satisfied that the electrode/voltage-preamplifier system was not a source of unwanted noise. The voltage preamplifier and electronics used in these diagnostic experiments were the same as those used in BMSR-2 for plant experiments.

For comparison, we also conducted identical tests with a low-noise *current* preamplifier (SRS Model SR570, 6-dB low-pass 10-Hz filter, sensitivity 100 nA/V). In this case the signal at 1.2 Hz did appear above the noise in the data of sensor C. For example, the 2-V experiment yielded a 3-pT signal along the $y$-axis, indicating that a current of over 20 nA was flowing in the current loop. Based on these results, we exclusively used the voltage preamplifier in our data run at PTB. As an additional security check, in all OPM experiments we ran one of the electrode wires over the background sensor D to monitor for possible spurious signals (none were detected).

Extended Data Fig. 5 shows the electric time traces used in the data analysis for Fig. 3 and Fig. 4 in the main text. The recorded APs are slightly variable in shape and exhibit certain artifacts, which is normal for surface-electrode measurements. In some time traces (e.g. Extended Data Fig. 5a) we observed the frequency of autonomous AP firing increasing over time, which may be explained as follows. Sufficient input energy is required to increase the cytosolic calcium level to threshold—once this threshold is reached, an AP is released. As the trap heats up in our setup, the stored cellular energy increases while the new energy which needs to be input for the next AP decreases, which could lead to an increase in AP firing frequency.

To characterize the performance of the QuSpin gradiometer system in the shielded room, we recorded the background in the room and performed frequency analysis. A typical noise spectrum is shown in Extended Data Fig. 6.

**Corresponding author**

Correspondence to Anne Fabricant.

**Extended data figures**

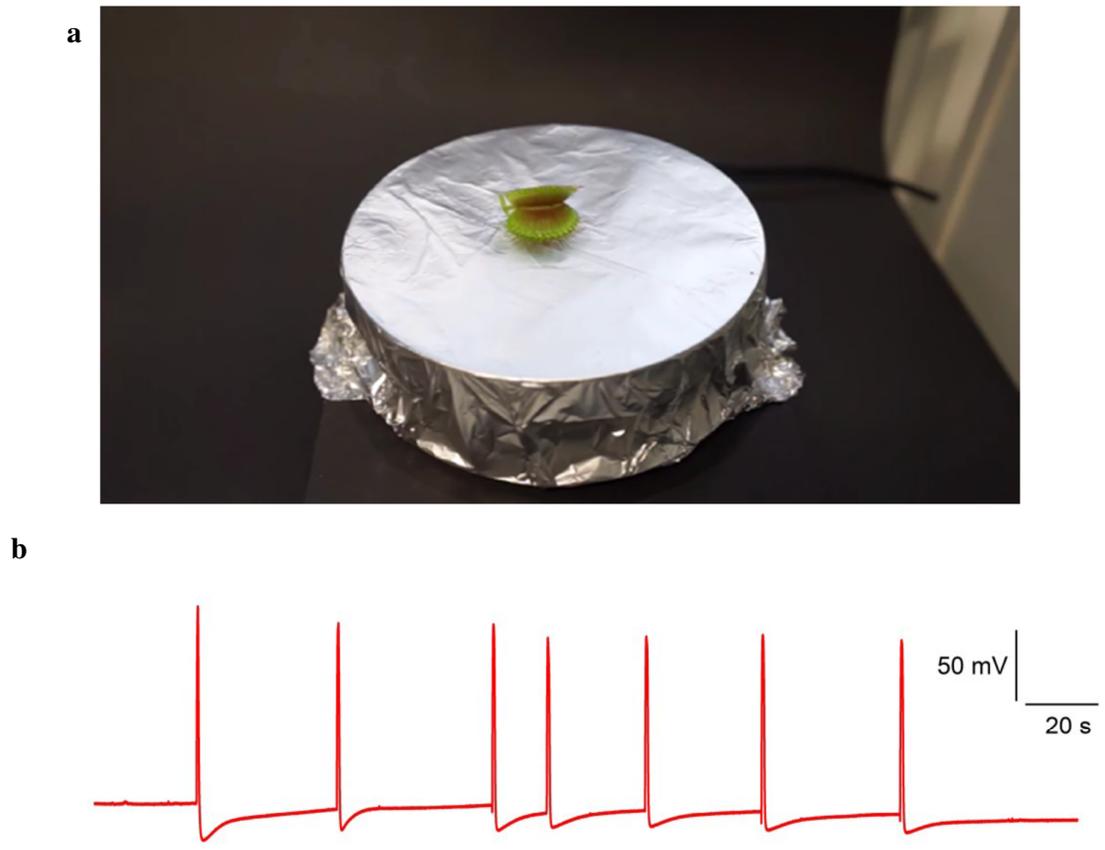

**Extended Data Fig. 1 | Spontaneous AP firing on a hot plate heated to 46°C. a**, Video (online) showing trap closure on the hot plate; playback speed is increased by a factor of 10. **b**, Surface-potential measurements of the heated trap, confirming that heat evokes APs.

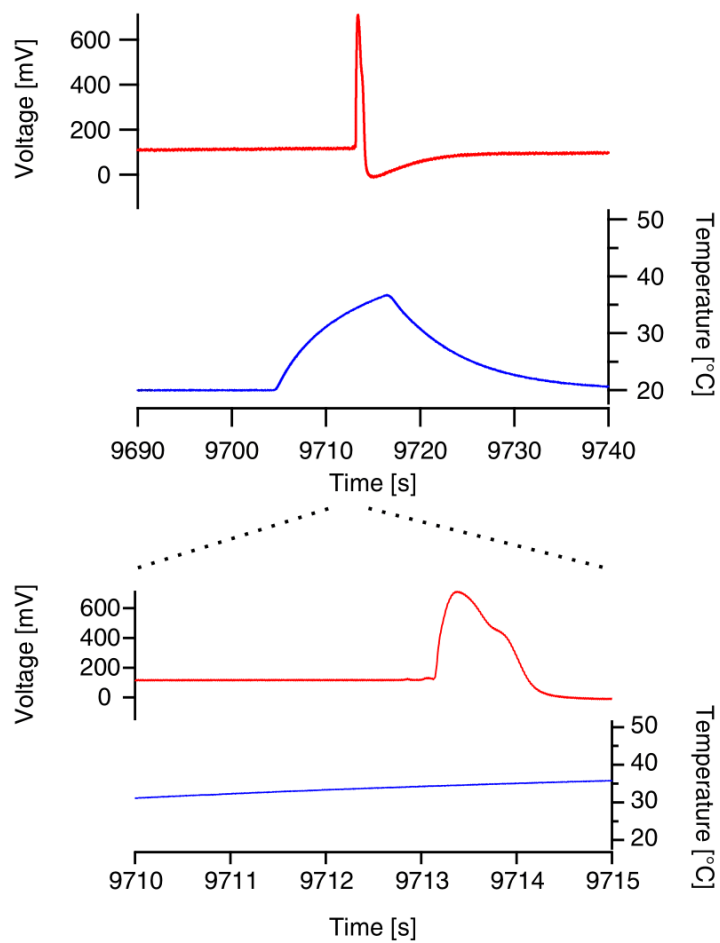

**Extended Data Fig. 2 | Comparison of measured surface potential and applied temperature.** Heat was applied via a Peltier element placed on the inner trap surface. An AP (red curve) occurred as the temperature (blue curve) increased from 20 to 45°C. The lower graph is a zoom-in on the time axis to define the temperature at which the AP occurred.

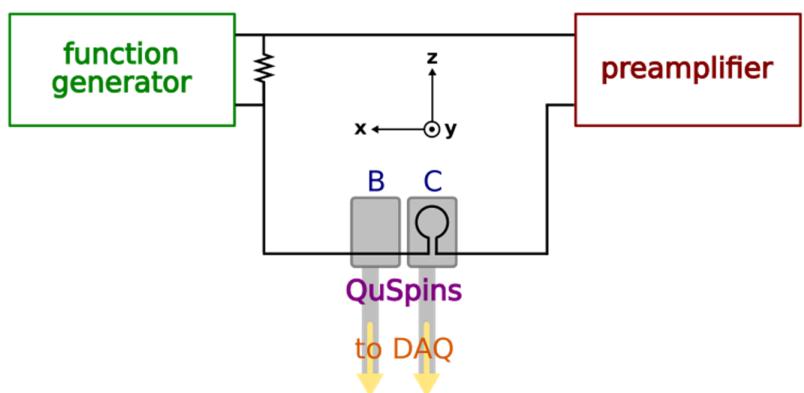

**Extended Data Fig. 3 | Circuit for testing the preamplifiers.** See Supplementary Information for details.

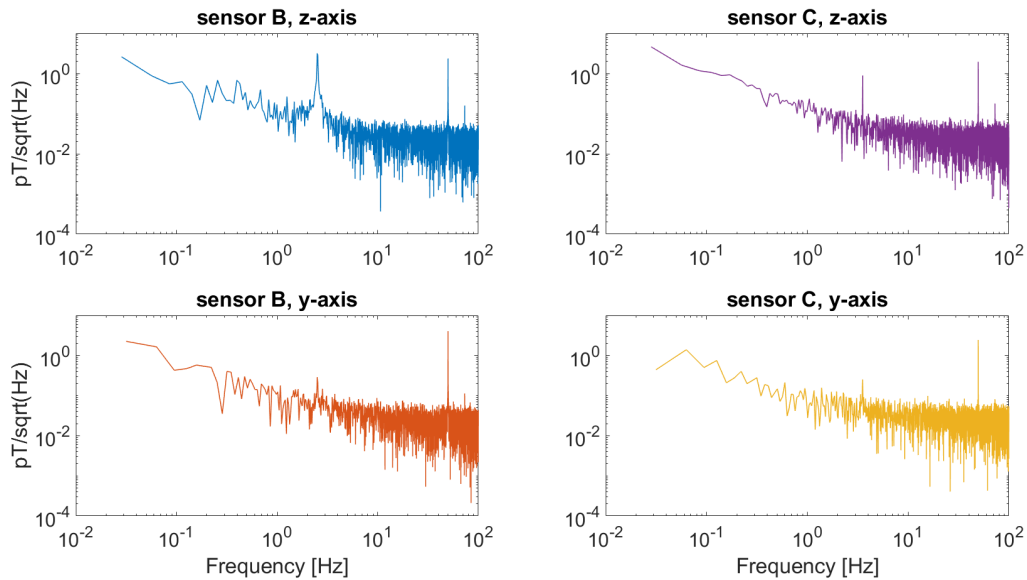

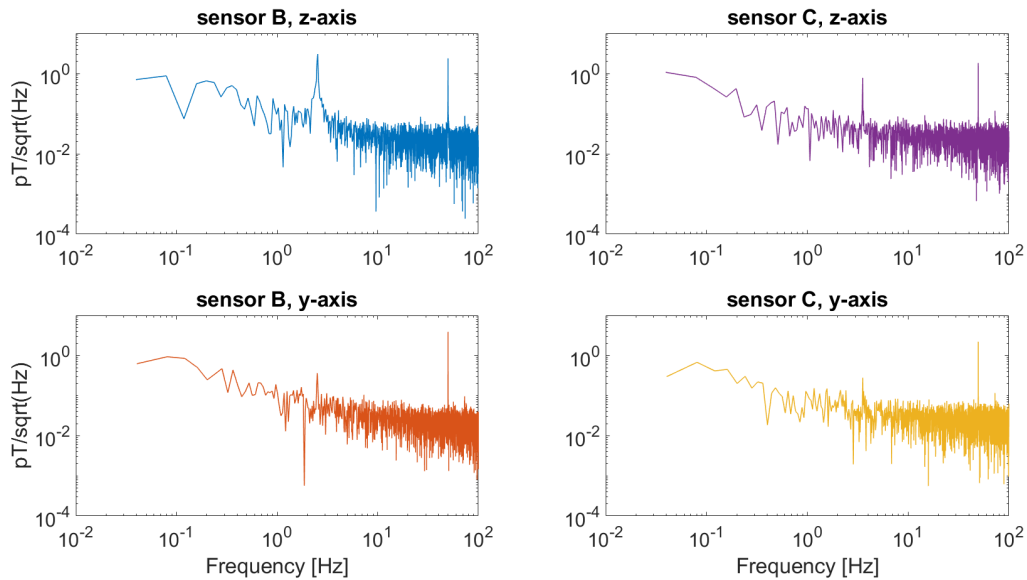

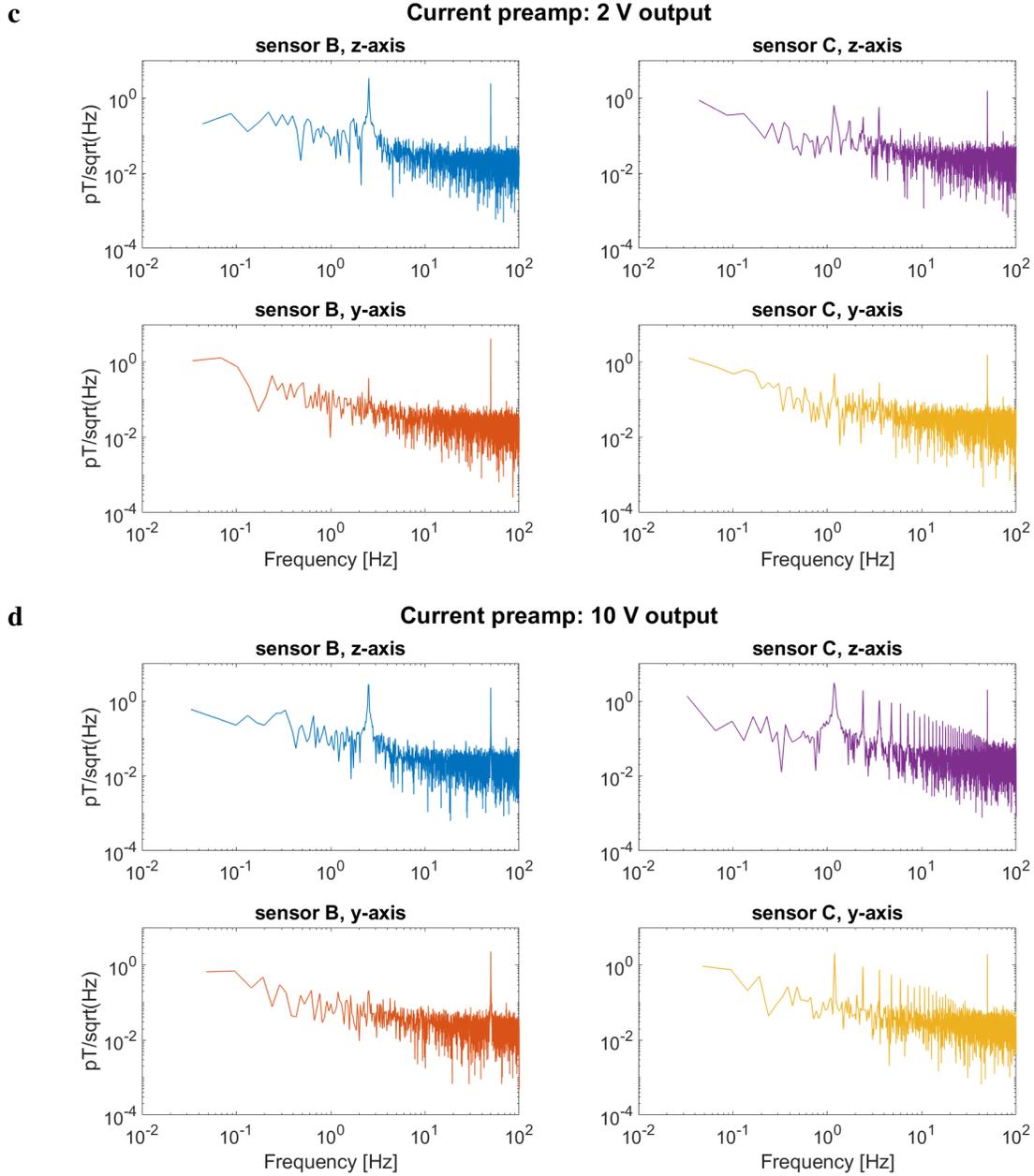

**Fig. S4 | Results of the preamplifier tests.** In addition to the 50-Hz line frequency, peaks due to lab background noise appear at 2.5 and 3.5 Hz. Signal from the current preamplifier appears in the data of sensor C (**c** and **d**), but this effect is not seen when the voltage preamplifier is used (**a** and **b**).

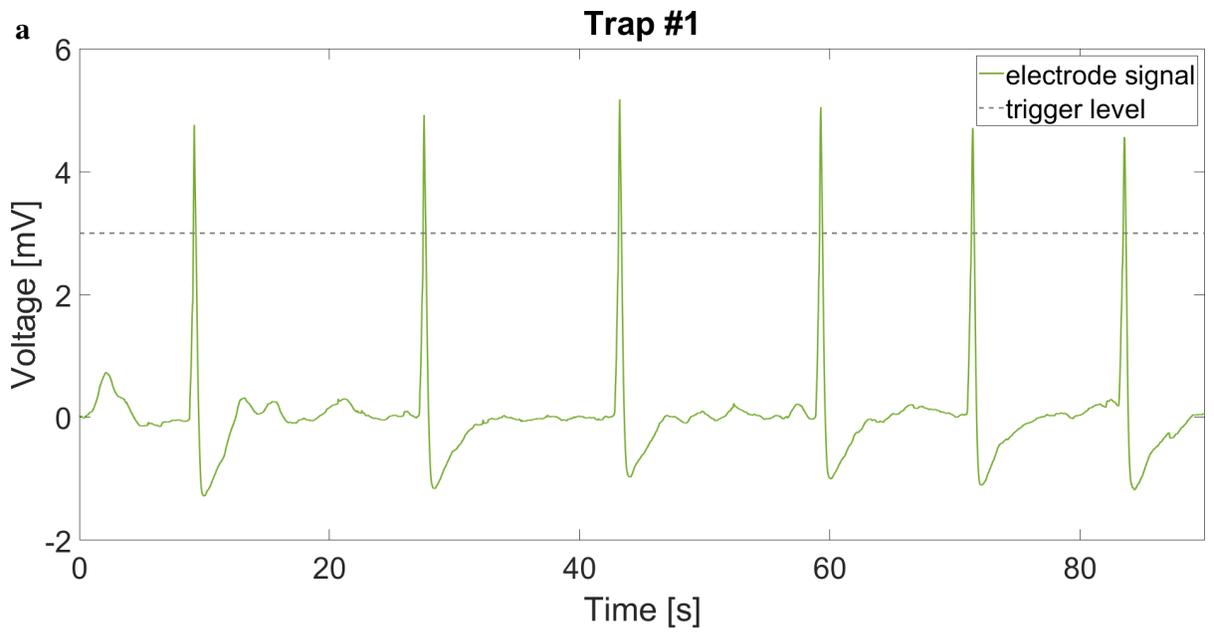

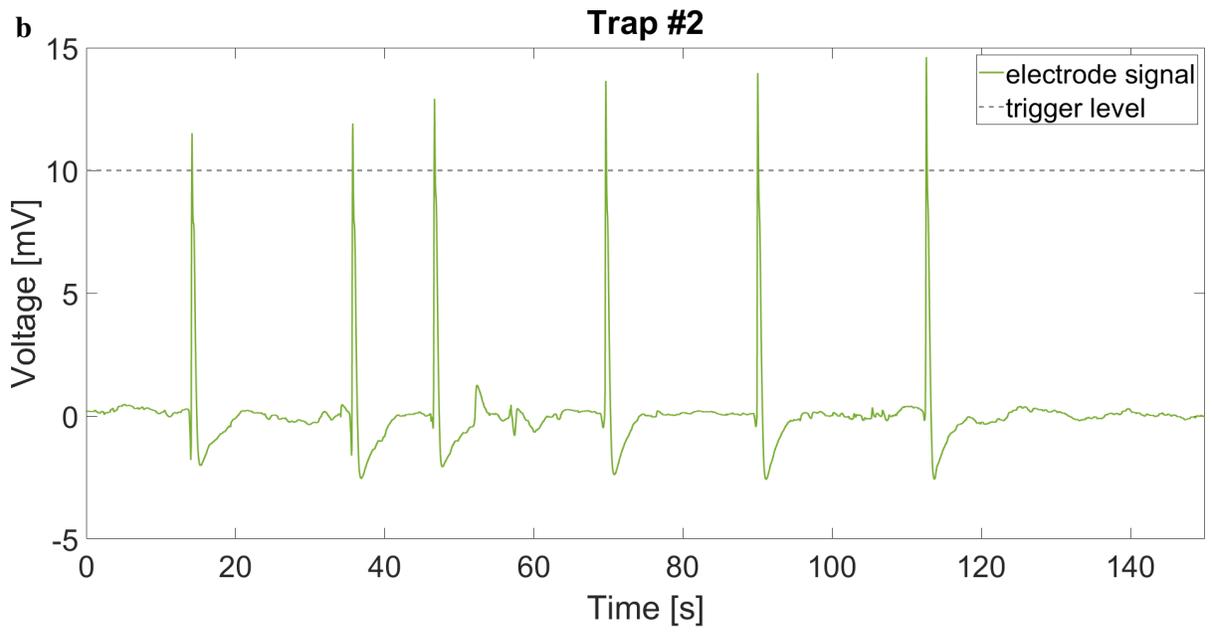

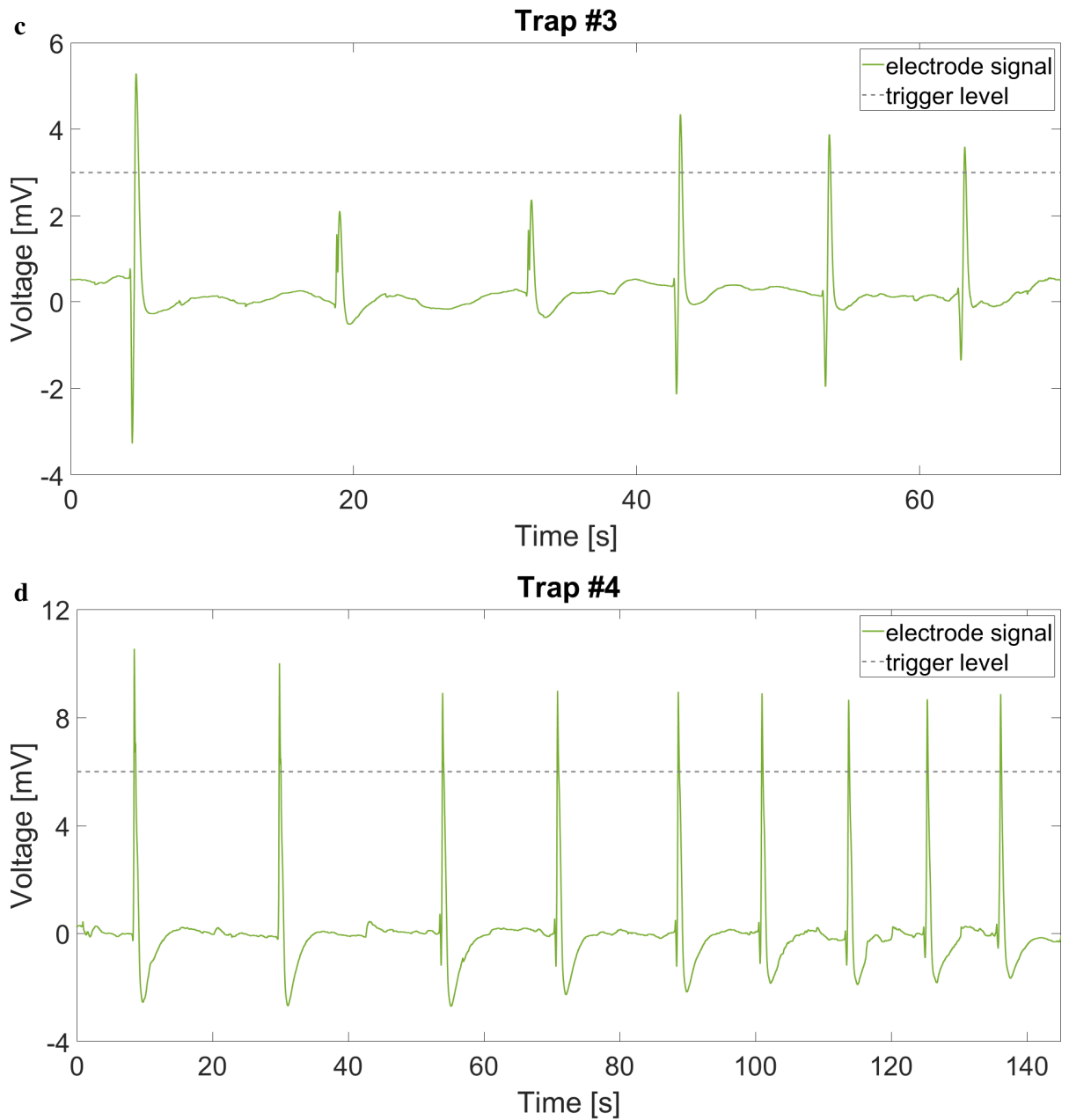

**Extended Data Fig. 5 | Electric time traces showing heat-induced action potentials in four separate experiments with different plant samples.** Corresponds to the data shown in Fig. 5 of the main text. The APs are used as a trigger so that we can perform averaging of the simultaneous magnetic data; the trigger level is indicated by the gray dashed line.

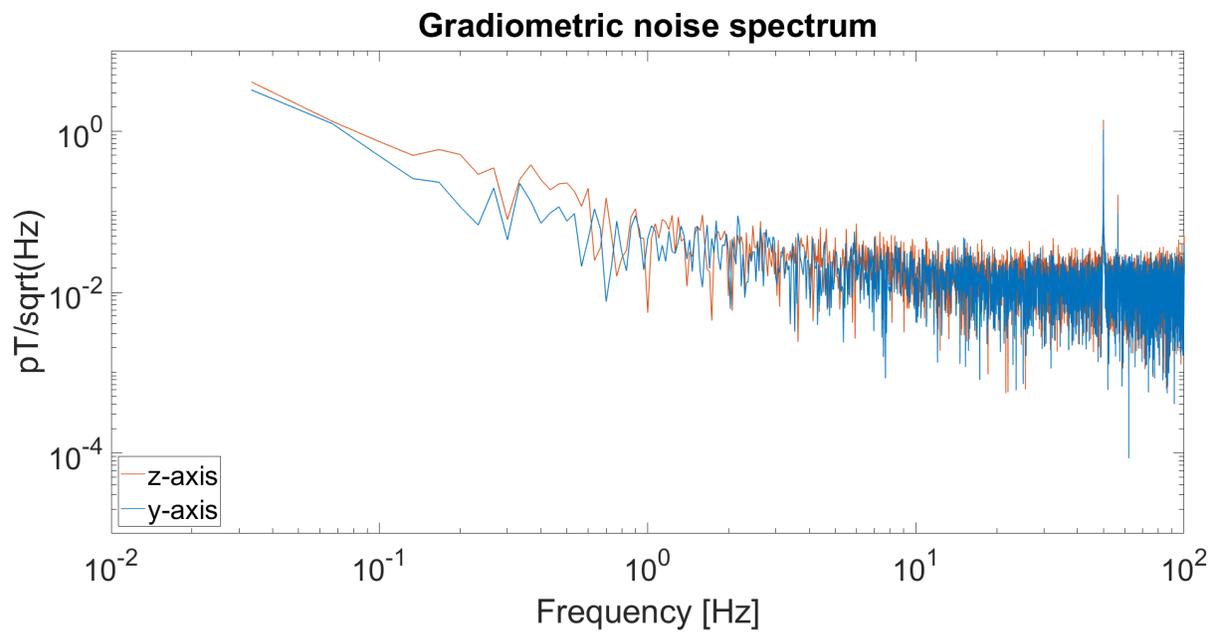

**Extended Data Fig. 6 | Typical noise floor of the gradiometer in the magnetically shielded room.** Obtained by recording a 30-s time trace prior to the start of an experiment.